\documentclass[aps,prl,twocolumn,floatfix,superscriptaddress,notitlepage]{revtex4-1}

\usepackage{amsmath}
\usepackage{amssymb}
\usepackage{graphicx}
\usepackage{float}
\usepackage[colorlinks, linkcolor=blue, citecolor=blue, urlcolor=blue, breaklinks=red]{hyperref}

\newcommand{\at}{\text{at}}

\begin{document}

\title{Collective Multi-mode Vacuum Rabi Splitting}
\author{W. Guerin}
\affiliation{Universit\'e C\^ote d'Azur, CNRS, Institut de Physique de Nice, France}
\author{T. S. do Espirito Santo}
\affiliation{Instituto de F\'{i}sica de S\~{a}o Carlos, Universidade de S\~{a}o Paulo - 13560-970 S\~{a}o Carlos, SP, Brazil}
\affiliation{IPCMS (UMR 7504) and ISIS (UMR 7006), Universit\'e de Strasbourg, CNRS, 67000 Strasbourg, France}
\author{P. Weiss}
\affiliation{Universit\'e C\^ote d'Azur, CNRS, Institut de Physique de Nice, France}
\author{A. Cipris}
\affiliation{Universit\'e C\^ote d'Azur, CNRS, Institut de Physique de Nice, France}
\author{J. Schachenmayer}
\affiliation{IPCMS (UMR 7504) and ISIS (UMR 7006), Universit\'e de Strasbourg, CNRS, 67000 Strasbourg, France}
\author{R. Kaiser}
\affiliation{Universit\'e C\^ote d'Azur, CNRS, Institut de Physique de Nice, France}
\author{R. Bachelard}
\affiliation{Departamento de F\'isica, Universidade Federal de S\~{a}o Carlos, Rod.~Washington Lu\'is, km 235 - SP-310, 13565-905 S\~{a}o Carlos, SP, Brazil}
\affiliation{Universit\'e C\^ote d'Azur, CNRS, Institut de Physique de Nice, France}

\begin{abstract}
We report the experimental observation of collective multi-mode vacuum Rabi splitting in free space. In contrast to optical cavities, the atoms couple to a continuum of modes, and the optical thickness of the cloud provides a measure of this coupling. The splitting, also referred as normal mode splitting, is monitored through the Rabi oscillations in the scattered intensity, and the results are fully explained by a linear-dispersion theory.
\end{abstract}

\date{\today}

\maketitle

Light scattering encompasses a broad range of phenomena, and its elementary brick can be found in the interaction of a vacuum mode with a single atom. From a fundamental point of view, the vacuum mode and the atom are two oscillators, whose coupling leads to hybrid modes with specific energies. In case the two oscillators possess the same natural frequency, the interaction lifts the degeneracy, and the new eigenenergies split by an amount proportional to the coupling. In the context of optical cavities, where the single mode hypothesis is best achieved, the phenomenon has been coined vacuum Rabi splitting~\cite{SanchezMondragon1983, Carmichael1986}. The effect has been successfully observed and utilized in a wide range of fields from cavity quantum electrodynamics in atomic physics~\cite{Thompson1992, Boca2004} to solid-state systems and chemistry~\cite{Yoshie2004, Reithmaier2004, Peter2005, Santhosh2016, Ask2019, Pockrand_Excito_1982, Lidzey_Strong_1998, Hutchison_Modify_2012, Shalabney_Cohere_2015, Agranovich_Hybrid_2011, Colombe2007}.

\begin{figure}
\includegraphics[width=.48\textwidth]{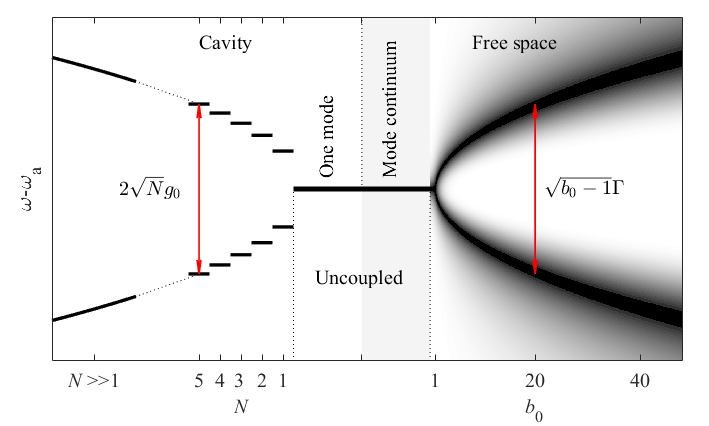}
\caption{Eigenenergies of an atomic cloud coupled to vacuum mode(s). Left: $N$ atoms coupled to a resonant single mode cavity; the mode splitting scales as $\sqrt{N}g_0$, with $g_0$ the single-atom coupling to the mode. Right: A cloud with resonant optical thickness $b_0$, coupled to the continuum of vacuum modes of free space; the mode splitting scales as $\sqrt{b_0-1}\Gamma$, with $\Gamma$ the single-atom decay rate in free space (see text for details).}
\label{fig:levels}
\end{figure}

Interestingly, before the strong coupling regime could be achieved experimentally with single atoms, the splitting was noted to be accessible experimentally in larger-volume cavities: If an ensemble of $N$ atoms are coupled to the cavity, the coupling strength to the vacuum mode is enhanced by a factor $\sqrt{N}$. This allowed for early observations in optical cavities~\cite{Orozco1987, Raizen1989, Brecha1995}, and it was later understood that a linear-dispersion theory could describe the phenomenon~\cite{Zhu1990, Yokoyama1997}. The fundamental difference between the single- and the many-atom case is that in the former case, the quantization of the electromagnetic mode becomes relevant~\cite{Brune1996} (see left part of Fig.\ref{fig:levels}).

Differently from high-finesse optical cavities, our three-dimensional world presents a continuum of vacuum modes, both in space and in frequency. Nevertheless, this multimode characteristics does not prevent the building up of collective modes. The first brick of collective scattering was laid down by Robert H. Dicke, when he showed that a collection of atoms, either in the small or the large volume limit, emit light at a ``superradiant'' rate~\cite{Dicke1954}. At first discussed in the quantum context of fully-inverted atoms, superradiant decay was later predicted in the limit of a single excitation~\cite{Scully_Direc_2006,Svidzinsky2008}, as confirmed by linear-optics measurements~\cite{Araujo_Super_2016, Roof_Obser_2016}. In the field of cooperative scattering, subradiance~\cite{Guerin_Subra_2016}, superflash~\cite{Kwong2014} and collective frequency shifts~\cite{Roof_Obser_2016,Rohlsberger2010,Keaveney2012,Okaba2014,Meir2014,Peyrot2018} contribute to the rich variety of observed phenomena.

The continuum of vacuum modes calls for a different modelling of the light-atom interaction in free space: An interpretation in terms of dipole-dipole interactions, obtained by tracing over the light degrees of freedom, is in general favored, as it allows addressing only the atomic degrees of freedom~\cite{Lehmberg_Radia1_1970}. Such a coupled-dipole approach was largely used to describe the cooperative phenomena described above. In particular, differently from optical cavities where the cooperativity parameter is $N$, the resonant optical thickness of the cloud was identified to play this role in free space for dilute clouds with a spatial extend larger than the optical wavelength~\cite{Guerin2016}.

\begin{figure*}
\includegraphics[width=1\textwidth]{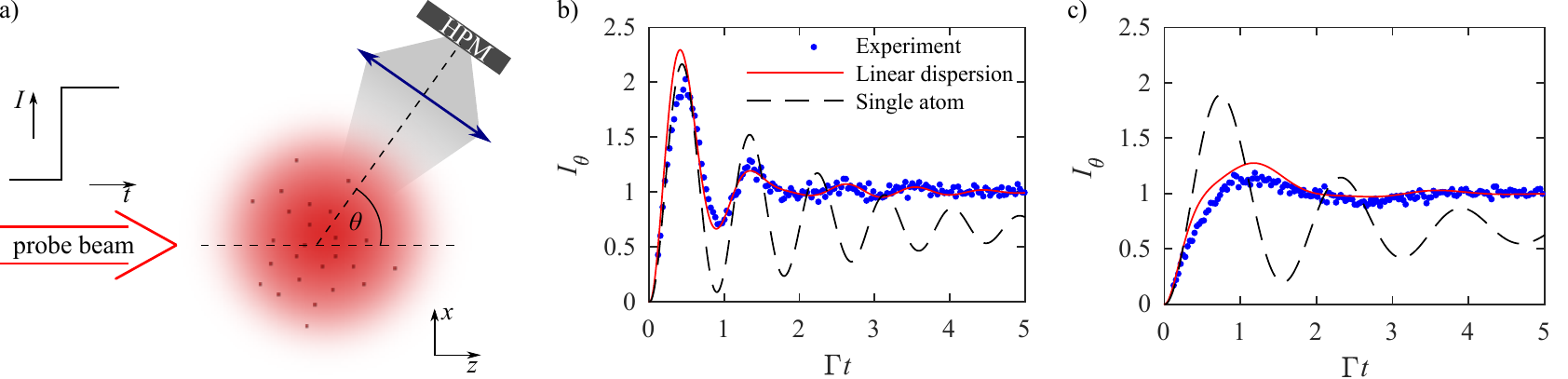}
\caption{(a) Sketch of the experiment: A monochromatic plane-wave laser suddenly illuminates a Gaussian cloud of two-level atoms, while the dynamics of the scattered intensity is measured at an angle of $\theta\approx 35^\circ$ from the axis of the incident beam by a hybrid photon detector (HPM). (b-c) Experimentally recorded intensity (dots) after the laser switch-on, normalized to 1 in the long-time limit, for different values of the optical thickness of the cloud and laser detunings [(b) $b_0\approx 8.1$ and $\Delta=-7\Gamma$; (c) $b_0\approx 20.3$ and $\Delta=-4\Gamma$]. 
The plain lines are obtained by a numerical simulation of the linear dispersion theory (see text), the dashed lines denote the single-atom response (not normalized, for visibility purposes).}
\label{fig:setup}
\end{figure*}

In this Letter, we report on the experimental signatures of collective multi-mode vacuum Rabi oscillations in free space, where the optical thickness acts as a measure of the coupling between the atomic cloud and the light modes (see right part of Fig.~\ref{fig:levels}). The splitting is monitored through the linear-optics Rabi oscillations of the cloud after an abrupt switch-on of the pump laser. Our measurements, realized over a set of driving frequencies and optical thicknesses, are in very good agreement with linear-dispersion theories for three-dimensional clouds.

Our experimental setup, which has been detailed in Ref.~\cite{Araujo_Super_2016}, is sketched in Fig.\ref{fig:setup}(a): A three-dimensional Gaussian cloud (rms width $R \approx 1\,\text{mm}$) of $N\approx 10^9$ randomly distributed $^{87}$Rb atoms is produced in a magneto-optical trap at a temperature $T\approx 100\mu K$.
The atoms are optically pumped to the $F = 2$ state, and driven on the $F=2\to F = 3$ transition (with wavelength $\lambda = 780.24\,\text{nm}$ and linewidth $\Gamma/2\pi = 6.07\,\text{MHz}$).
The cloud is homogeneously illuminated by a linearly polarized laser beam (waist $w\approx5.7\,\text{mm}$, detuning $\Delta=\omega_\mathrm{L}-\omega_\mathrm{a}$ from the atomic transition) propagating along the $z$-axis. A series of pulses with 10-90\% rise time of about 6\,ns, short compared to the lifetime of the excited state $\tau_\at = \Gamma^{-1} = 26.2\,$ns, are produced by acousto- and electro-optical modulators. During the series of pulses, the atomic cloud expands ballistically, which allows us to probe different on-resonance optical depths, defined as $b_0 = \sigma_0^\mathrm{Rb}\int \rho(0,0,z)dz$, with $\sigma_0^\mathrm{Rb}$ the resonant Rubidium atomic cross-section. 
The light intensity is adjusted to keep a constant saturation parameter $s=2\Omega_0^2/(\Gamma^2+4\Delta^2) \simeq (2.2 \pm 0.6) \times 10^{-2}$, with $\Omega_0$ the Rabi frequency of the laser. The time-dependent scattered light intensity is recorded by a photon detector in the far field at an angle of $\theta \approx 35^\circ$ from the laser axis. The finite rise time of the laser field, as well as small spurious overshoots, are accounted for by first dividing the recorded signal by the switch-on temporal profile of the laser alone (recorded with the same cycle and same detector with light scattered on white paper), in order to focus on the atomic dynamics. 

Typical examples of intensity signal from the experiment are presented in Figs.~\ref{fig:setup}(b) and (c). The radiation emitted by the cloud of cold atoms exhibits Rabi oscillations, whose frequency are smaller than the single-atom one: For a single atom, one indeed expects an oscillation of the excited state population with the generalized Rabi frequency $\Omega_R = \sqrt{\Delta^2+\Omega_0^2}\approx|\Delta|$ in the linear optics regime considered here. As we shall now show, the deviation from this single-atom oscillation frequency finds its origin in the large resonant optical thickness of the cloud, which considerably enhances its dispersive features, and provides a direct measurement of the coupling strength between the cloud and the vacuum modes. 

Differently from optical cavities, in free space the three-dimensional continuum of vacuum modes forces us to adopt a space-dependent theory in order to investigate the dispersion properties of the cloud.  In our setup, let us first consider the propagation of a wave scattered off an atom at position $\mathbf{r}$. The wave of frequency $\omega$ moves from an initial position $\mathbf{r}_i = \mathbf{r} - w \hat z$ ($w\to \infty$) to the atom at $\mathbf{r}$ where it is scattered, and then to the detector positioned at  $\mathbf{r}_f = \mathbf{r} + w\hat \theta$, where $\hat z$ and $\hat \theta$ denote the unit vectors of the $z$-axis and the detector direction with respect to the atom, respectively. In the dilute regime with atoms being distributed in the cloud by a density distribution function, $\rho(\mathbf{r})$, and being uniformly illuminated by the laser, this process can be described by the following transfer function for an atom in $\mathbf{r}$~\cite{Kuraptsev2017}:
\begin{equation}\label{eq:Tom}
t_{\mathbf{r},\theta}(\omega)= \frac{\exp\left(-\frac{1}{2}\frac{b_{\mathbf{r}-w\hat{z}}^{\mathbf{r}}+b_\mathbf{r}^{\mathbf{r}+w\hat\theta}}{1 -2i(\omega-\omega_a)/\Gamma}\right)}{1-2i(\omega-\omega_a)/\Gamma}
\end{equation}
Here, 
$b_\mathbf{r}^{\mathbf{r}'}=\sigma_\textrm{sc}\int_{\mathbf{r}}^{\mathbf{r}'}d\mathbf{r}\rho(\mathbf{r})$ denotes the optical thickness for a light ray propagating from $\mathbf{r}$ to $\mathbf{r}'$. The expression assumes $w\gg R$, which corresponds to a detector in the far-field.
Although it neglects disorder in single experimental realizations, this approach captures well collective phenomena such as superradiance~\cite{Kuraptsev2017}. In Fig.\ref{fig:setup}(b), one can observe that it captures very well the damped oscillations of the radiated intensity, but also the beating that is absent from the single-atom response.

In the simplified case of an infinite slab illuminated by a plane-wave, the forward-scattering response ($\theta=0$) is obtained by integrating the local response over the cloud, $t_0(\omega)=\int d\mathbf{r}\rho(\mathbf{r})t_{\mathbf{r},\theta}(\omega)$, which leads to 
\begin{equation}
t_0(\omega)\sim \frac{\exp\left(-\frac{b_0}{2}\left[1-\frac{2i(\omega-\omega_a)}{\Gamma}\right]^{-1}\right)}{1-2i(\omega-\omega_a)/\Gamma},
\end{equation}
where $b_0$ is the resonant optical thickness of the slab.
By analyzing the response function $t_0(\omega)$, one observes that for $b_0>1$, the cloud response splits into two symmetric resonances of frequency
\begin{equation}
    \omega_\pm=\omega_a\pm \frac{\sqrt{b_0-1}}{2}\Gamma.\label{eq:split1d}
\end{equation}
This behaviour is illustrated in the right part of Fig.~\ref{fig:levels}:
The role of the coupling strength between the cloud and the vacuum modes, quantified by the resonance splitting, is thus here assumed by the resonant optical thickness. 
\begin{figure}
\includegraphics[width=.45\textwidth]{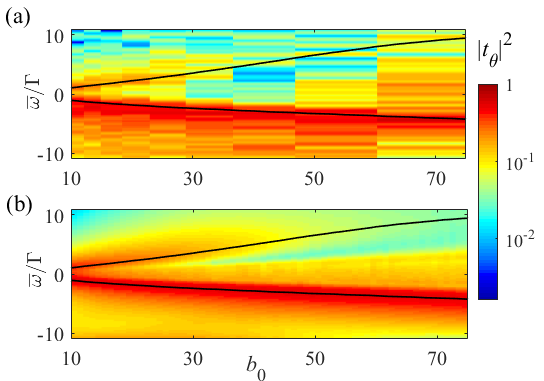}
\includegraphics[width=.42\textwidth]{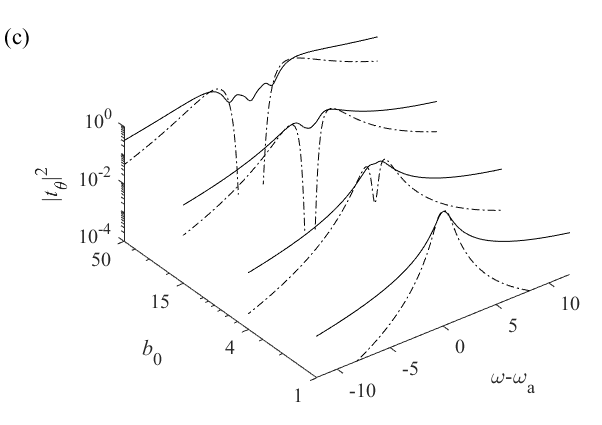}
\caption{Frequency response of the cloud to the switch-on of the laser, at an angle $\theta=35^\circ$. $|t_\theta(\omega)|^2$ is computed from the Fourier transform of the intensity (a) from the experiment and (b) from the linear-dispersion theory. The black lines denote, in both figures, the maxima derived from the linear-dispersion approach. $\overline{\omega}=\omega_\mathrm{opt}-\omega_\mathrm{L}$, i.e., it corresponds to the difference between the optical frequency and the laser frequency (here detuned by $\Delta=-12\Gamma$). (c) Frequency response of the cloud for increasing optical thicknesses, for a slab and in the forward
direction (dash-dotted curves), and for a spherical cloud
with a Gaussian density and for light at an angle
$\theta = 35^\circ$ (plain lines).}
\label{fig:fig3}
\end{figure}

For our spherical Gaussian cloud, we call the observation angle $\hat\theta$ and normalize the distances as $\mathbf{r}\to\mathbf{r}/R$:
\begin{eqnarray}
b_{\mathbf{r}-w\hat{z}}^{\mathbf{r}} &=& \frac{b_0}{2}e^{-\frac{r^2-(\mathbf{r}.\hat{z})^2}{2}}\left[1+\textrm{Erf}\left(\frac{\mathbf{r}.\hat{z}}{\sqrt{2}}\right)\right]
\\ b_\mathbf{r}^{\mathbf{r}+w\hat{\theta}} &=& \frac{b_0}{2}e^{-\frac{r^2-(\mathbf{r}.\hat{\theta})^2}{2}}\left[1-\textrm{Erf}\left(\frac{\mathbf{r}.\hat{\theta}}{\sqrt{2}}\right)\right],\nonumber
\end{eqnarray}
where the limit $w\to\infty$ is used.
The cloud response for the Gaussian sphere is then computed numerically using Eq.~\eqref{eq:Tom}, which in turn allows to obtain the intensity dynamics in the temporal domain: After multiplying it by the Fourier transform of the pump beam temporal profile (a Heaviside function), the local frequency response of the atoms is converted into a temporal response by computing its inverse Fourier-transform; the obtained intensity is then integrated over all the cloud~\cite{Kuraptsev2017,Chalony2011}. The temporal curves presented in Fig.~\ref{fig:setup} were obtained this way. We have also checked that the microscopic coupled dipole model discussed in the introduction~\cite{Lehmberg_Radia1_1970,Svidzinsky_Coope_2010} provides temporal signals in excellent agreement with the linear-dispersion approach~\eqref{eq:Tom}~\cite{EspiritoSanto19}. An important difference, though, is that these microscopic simulations are limited to thousands of particles, so they address much smaller systems. 

The measurement of the splitting requires the oscillations to be faster than the decay rate~\cite{Kaluzny1983}. Close to resonance, Close to resonance, we
observe that the oscillations vanish, see
Fig.~\ref{fig:setup}(c), which we attribute to occurrence of multiple scattering ($b(\Delta)=b_0/(1+4\Delta^2/\Gamma^2)>1$)~\cite{Araujo_Super_2016}. Out of resonance (such that $b(\Delta)<1$), the detuning is chosen such that the single-atom generalized Rabi frequency $\Omega_R\approx|\Delta|$ be larger than the superradiant decay rate~\cite{Araujo_Super_2016}, which allows to monitor the deviation from the single-atom oscillations, see Fig.~\ref{fig:setup}(b). 
Note that while the experimental curves are obtained from averaging the signal over thousands of runs, the linear-dispersion effective-medium approach provides a signal that is naturally fluctuation-free.

Let us now discuss in more details the splitting in frequency space: In Fig.~\ref{fig:fig3}(a,b), the Fourier transform of the intensity signal is presented, for the experimental data and the linear-dispersion simulations. We first note that while the splitting is present, the lower branch $\omega_-$ is much more visible, especially at high $b_0$. This is due to the fact that the system is pumped with a negative detuning ($\Delta=-12\Gamma$ in this case), so the laser couples more strongly to this branch. The increasing optical thickness makes this lower branch even closer to the laser frequency, and the upper branch $\omega_+$ even farther, which results in an increasing imbalance between the branches. A pumping at $\Delta\gg \omega_\pm$ would allow to populate almost equally the two branches, yet at the price of a weaker radiated intensity.

As discussed above (see Eq.\eqref{eq:split1d}), the one-dimensional geometry of the slab provides a simple scaling $\Delta\omega\sim\sqrt{b_0-1}$ for the splitting, see Eq.\eqref{eq:split1d} and Fig.\ref{fig:levels}. The three-dimensional Gaussian cloud used in our experiment, with an observation angle at $\theta=35^\circ$ and a beam larger than the cloud, leads to the coupling of the atoms to a larger family of vacuum modes. This is illustrated in Fig.\ref{fig:fig3}(c), where the difference of the splitting process for the slab and the Gaussian sphere is presented. The slab is characterized by a strong gap between the two resonances, whereas the Gaussian sphere presents a rather shallow dip, with even the emergence of secondary resonances at higher $b_0$. Furthermore, the decay of the tails of the frequency response are much slower for the Gaussian sphere. One explanation for these differences is that in an ideal (i.e., infinite) slab, all incoming rays see the same optical thickness. Differently, in a Gaussian sphere rays outside the $z$-axis go through the medium with a different optical thickness; the presence of this variety of optical thicknesses for a three-dimensional cloud (an effect reinforced by the observation angle $\theta>0$) leads to a broadening of the cloud dispersive response. More generally, in a single-mode cavity, the system possesses only two modes (plus $N-1$ degenerate dark states), and the large $N$ limit makes the quantization of the photons irrelevant; but in free space, a single atom couples to a continuum of light modes, already leading to a broad response.

\begin{figure}
\includegraphics[width=.45\textwidth]{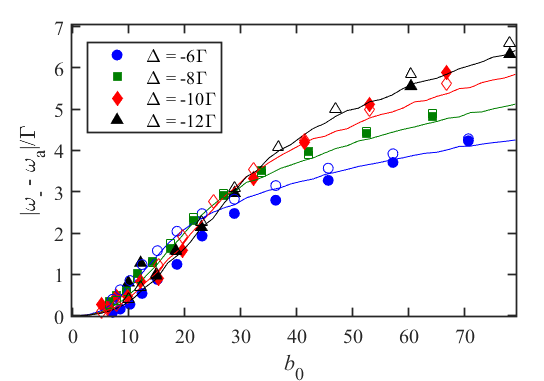}
\caption{Splitting amplitude extracted from the experiment (full symbols), from the coupled dipole simulations (empty symbols) and from the linear dispersion theory (lines), for different detunings and optical thicknesses.}
\label{fig:fig4}
\end{figure}

The shift of the dominant branch is then systematically deduced from the experimental data by fitting it to a single-dipole function $I_\theta(t)=|1-\exp[(i(\Delta+\omega_--\Gamma_\mathrm{SR}/2)t]|^2$, with $\omega_-$ and $\Gamma_\mathrm{SR}$ as fitting parameters. This procedure yields less fluctuations than the Fourier transform. The results are presented in Fig.~\ref{fig:fig4}, for a set of detunings and resonant optical thicknesses. They are compared to the linear dispersion theory results and to the coupled dipole model using the same procedure. All three approaches present results in very good agreement. The different values of detuning used, as well as the different system sizes simulated (in particular for the coupled dipole approach) highlight the role of the resonant optical thickness $b_0$ as a measure of the collective coupling of the atomic cloud to the vacuum modes in free space. We note that for larger values of $b_0$, an increasing detuning $|\Delta|$ is necessary to obtain a splitting that does not depend on the detuning, an effect which we attribute to multiple scattering.

We stress that the collective splitting discussed here is fundamentally different from the ``collective Lamb shift'' (CLS) reported in atomic systems~\cite{Roof_Obser_2016,Rohlsberger2010,Keaveney2012,Okaba2014,Meir2014,Peyrot2018}. In the larger sample limit ($R\gg\lambda$), the CLS scales with the atomic density~\cite{Friedberg_Frequ_1973,Friedberg_Analy_2010,Manassah2012}, whereas the present experiment was realized using dilute clouds ($\rho/k^3\sim0.01$). 
In particular, we point out that the splitting  for the slab (see Eq.~\eqref{eq:split1d}) depends only on $b_0$, and does not present any explicit dependence on the atomic density, which clearly differentiates it from the CLS.
We note that the oscillations which emerge from the light-atom coupling, sometimes called ``ringing'', has stimulated several experimental and theoretical works~\cite{Kaluzny1983,Pavolini1985,Cummings1986}. Here, within the context of free space linear-optics, we propose a unified picture of the macroscopic coupling between an atomic cloud and vacuum modes.

As a final remark, it is interesting to note the connection between steady-state frequency-resolved spectroscopy and time-dependent spectroscopy. In cavity spectroscopy, a well-known technique called ring-down spectroscopy~\cite{Herbelin1980} has been developed, where a photon bullet picture, which neglects interference effects, can describe the observed phenomena, whereas cavity transmission experiments involves interference effects between multiple reflections inside the cavity. For our mirrorless configuration, steady-state experiments sensitive to the frequency-dependent scattered intensity $T(\omega)=|t(\omega)|^2$ have been presented in Ref.~\cite{Labeyrie2004}. In contrast, the data presented in this work have been obtained in time-dependent experiments and depend on the Fourier transform of $t(\omega)$. Similar to the situation of cavity spectroscopy, we expect the sensitivity to intensity and phase fluctuations to scale differently using these different protocols, a feature which might be exploited when considering fluctuations or dephasing mechanisms.

In conclusion, we have reported on the experimental observation of the collective multi-mode vacuum Rabi splitting in free space, by monitoring the linear-optics Rabi oscillations of the scattered intensity after an abrupt switch-on of the pump laser. The scaling of the splitting with the resonant optical thickness shows that the latter is a measure of the coupling between the atomic clouds and the three-dimensional continuum of vacuum modes in free space.

\begin{acknowledgments}
We thank Ivor Kre\v{s}ic and Michelle Ara\'ujo for their contribution in setting up the fast switch-on system and Luis Orozco for fruitful discussions. Part of this work was performed in the framework of the European Training Network ColOpt, which is funded by the European Union (EU) Horizon 2020 programme under the Marie Sklodowska-Curie action, grant agreement No. 721465. R. B. and T. S. E. S. benefited from Grants from S\~ao Paulo Research Foundation (FAPESP) (Grants Nos. 2018/01447-2, 2018/15554-5 and 2019/02071-9) and from the National Council for Scientific and Technological Development (CNPq) Grant Nos. 302981/2017-9 and 409946/2018-4. R. B. and R. K. received support from project CAPES-COFECUB (Ph879-17/CAPES 88887.130197/2017-01). J.S.~is supported by the French National Research Agency (ANR) through the Programme d'Investissement d'Avenir under contract ANR-11-LABX-0058\_NIE within the Investissement d'Avenir program ANR-10-IDEX-0002-02 and by QuantERA - Project ``RouTe''. P.W.~is supported by the Deutsche Forschungsgemeinschaft (grant WE 6356/1-1). The Titan X Pascal used for this research was donated by the NVIDIA Corporation.
Research carried out using the computational resources of the Center for Mathematical Sciences Applied to Industry (CeMEAI) funded by FAPESP (grant 2013/07375-0).
\end{acknowledgments}

\bibliography{CVRS}

\end{document}